\documentclass[journal=nalefd,manuscript=letter,layout=twocolumn]{achemso}

\usepackage[version=3]{mhchem} % Formula subscripts using \ce{}

\pdfoutput=1

\usepackage[english]{babel}
\usepackage[utf8]{inputenc}
\usepackage{newtxtext, newtxmath}
\usepackage[scaled=0.8]{beramono}
\usepackage[T1]{fontenc}
\usepackage{breqn}

\usepackage{acronym}
\usepackage{bm}
\usepackage{dsfont}
\usepackage{graphicx}
\usepackage{mathtools}
\usepackage{microtype}
\usepackage{tikz}

\usetikzlibrary{decorations.markings}
\usetikzlibrary{decorations.pathmorphing}

\definecolor{mauve}{RGB}{94, 60, 153}

\usepackage[colorlinks, allcolors=mauve]{hyperref}
\usepackage[figure, figure*]{hypcap}

\def\Zagreb{Centre for Advanced Laser Techniques, Institute of Physics, 10000 Zagreb, Croatia}

\def\Kiel{Institut f\"ur Theoretische Physik und Astrophysik, Christian-Albrechts-Universit\"at zu Kiel, 24098 Kiel, Germany}

\def\Kiell{Kiel Nano, Surface and Interface Science KiNSIS, 24118 Kiel, Germany}

\author{Nina Girotto}
\affiliation\Zagreb

\author{Fabio Caruso}
\affiliation\Kiel
\altaffiliation\Kiell

\author{Dino Novko}
\email{dino.novko@gmail.com}
\affiliation\Zagreb

\title{Ultrafast nonadiabatic phonon renormalization in photoexcited single-layer MoS$_2$}

\begin{document}

\sloppy

\begin{abstract}
Comprehending nonequilibrium electron-phonon dynamics at the microscopic level and at the short time scales is one of the main goals in condensed matter physics. Effective temperature models and time-dependent Boltzmann equations are standard techniques for exploring and understanding nonequilibrium state and the corresponding scattering channels. However, these methods consider only the time evolution of carrier occupation function, while the self-consistent phonon dressing in each time instant coming from the nonequilibrium population is ignored, which makes them less suitable for studying ultrafast phenomena where softening of the phonon modes plays an active role. Here, we combine \emph{ab-initio} time-dependent Boltzmann equations and many-body phonon self-energy calculations to investigate the full momentum- and mode-resolved nonadiabatic phonon renormalization picture in the MoS$_2$ monolayer under nonequilibrium conditions. Our results show that the nonequilibrium state of photoexcited MoS$_2$ is governed by multi-valley topology of valence and conduction bands that brings about characteristic anisotropic electron-phonon thermalization paths and the corresponding phonon renormalization of strongly-coupled modes around high-symmetry points of the Brillouin zone. As the carrier population is thermalized towards its equilibrium state, we track in time the evolution of the remarkable phonon anomalies induced by nonequilibrium and the overall enhancement of the phonon relaxation rates. This work shows potential guidelines to tailor the electron-phonon relaxation channels and control the phonon dynamics under extreme photoexcited conditions.
\end{abstract}

\vspace{10mm}

\section{Introduction}

Recent advancements of ultrafast spectroscopy techniques have opened many avenues for controlling and understanding fundamental interactions in quantum materials\,\cite{giannetti16,torre21}. Due to ultrashort duration, usually below the characteristic thermalization timescale, the corresponding laser sources are not only able to probe and disentangle the relaxation pathways of electrons, phonons, spin, and other degrees of freedom\,\cite{petek97,bauer15}, but can reveal new physical phenomena and phases of matter beyond the thermodynamical equilibrium\,\cite{giannetti16}. Namely, photo-induced nonequilibrium carrier distribution and the accompanying modifications of the potential energy landscape was shown to promote nonlinear lattice control\,\cite{forst11,pomarico17,sentef17}, elevate or quench existing superconducting phase\,\cite{fausti11,zhang14,mitrano16,kennes17,giusti19}, alter the transition temperature\,\cite{schmitt08,rohwer11} or dimensionality\,\cite{duan21,cheng22} of the known charge-density-wave (CDW) order\, or even induce new ordered phases of matter\,\cite{stojchevska14,haupt16,kogar20,liu21,maklar21,zhang22,huang22}, as well as switch ferroelectric\,\cite{mankowsky17,nova19,li19,shin22,krapivin22,song23,fechner23} and ferromagnetic\,\cite{disa23} properties.

Electron-phonon coupling (EPC) plays a crucial role in the aforesaid ultrafast phenomena\,\cite{carbone08,giannetti16,kemper17,baldini20} and thus it is of utmost importance to master and comprehend microscopic channels ruling phonon dynamics in extreme nonequilibrium conditions. Complementary to the time-resolved photoemission methods that provide an important access to the electron-hole thermalization process\,\cite{petek97,bauer15,cabo15,beyer19,na19,duvel22} and electronic structure changes\,\cite{pogna16,wood20}, there are several ultrafast techniques, such as ultrafast electron diffraction scattering\,\cite{waldecker17,lin17,stern18,maldonado20,otto21,seiler21,krishnamoorthy19,britt22}, coherent phonon spectroscopy\,\cite{ishioka08,gerber17,hein20,sidiropoulos21,bae22}, and time-resolved Raman spectroscopy\,\cite{yan09,wu12,katsiaounis21,ferrante22}, that can precisely track the phonon relaxation channels following the photoexcitation and corresponding EPC strength\,\cite{gerber17,hein20}. For instance, ultrafast electron diffraction had uncovered highly anisotropic non-thermal phonon relaxation in black phosphorus\,\cite{seiler21}, and mapped momentum-resolved electron-phonon scattering channels and strengths in various transition metal dichalcogenides (TMDs)\,\cite{waldecker17,krishnamoorthy19,otto21,britt22}. Intriguingly, these methods are able to analyze photo-induced phonon frequency modifications and uncover the relevant microscopic processes, as it was done, for example, for zone-center strongly-coupled $E_{2g}$ optical mode in graphite with coherent phonon\,\cite{ishioka08} and time-resolved Raman spectroscopies\,\cite{yan09}, as well as for the amplitude CDW mode in TiSe$_2$ by means of ultrafast electron diffraction\,\cite{otto21}. In combination with other time-resolved spectroscopy approaches, the latter technique allowed to pinpoint the phonon modes that play an active role in unconventional superconductivity of FeSe thin films on SrTiO$_3$, and to extract the correlation-induced EPC constants\,\cite{gerber17}. Further, recent study on graphite reported dynamics of coherent vibrations for both zone-center and zone-edge phonon modes with unprecedented energy and time resolution allowed by attosecond core-level spectroscopy\,\cite{sidiropoulos21}.

In order to obtain a complete insight into the ultrafast phonon dynamics and unveil the corresponding electron-phonon scattering paths, the above studies need to be complemented with microscopic theoretical methods, preferably based on quantitative description of nonequilibrium state beyond simple phenomenological approaches. Probably the most widespread theoretical description of electron-lattice energy flow is the two temperature model (TTM) and its extensions\,\cite{allen87,perfetti07,koopmans09,johannsen13,waldecker16,maldonado17,caruso22}, where it is assumed that electron and lattice subsystems are already thermalized. It is commonly used in its phenomenological form to supplement the experimental observations of the heat transfer\,\cite{perfetti07,koopmans09,johannsen13,majchrzak21}, and was further useful to address the hot phonon dynamics and phonon bottleneck in materials with highly anisotropic EPC, such as graphene\,\cite{johannsen13,caruso20,caruso22}, graphite\,\cite{ishida11}, MgB$_2$\,\cite{novko20b,cappelluti22}, MoS$_2$\,\cite{han23}, and lead halid perovskites\,\cite{chan21}. It was also used as a basis to study laser-induced energy renormalization of hot phonons in MgB$_2$ as a function of time\,\cite{novko20b}.

However, the TTM is likely to fail in describing femtosecond lattice dynamics below the electron thermalization time, where full information of energy-momentum phase space in nonequilibrium state is required\,\cite{caruso22}. A considerable improvement in describing nonequilibrium scattering events is met with time-dependent Boltzmann equations (TDBE), especially with its first-principles implementations\,\cite{jhalani17,sadasivam17,seiler21,caruso21,tong21,britt22}. When both electron-phonon and phonon-phonon scattering channels are included, the TDBE can accurately describe energy-, momentum-, and time-dependent modifications to electron and phonon population in both sub-picosecond and picosecond regimes\,\cite{caruso17}. Despite these strengths, common TDBE studies do not account for the self-consistent renormalization of phonon energies in each time instant coming from the updated carrier populations\,\cite{caruso22}, and are, therefore, not fully suited for exploration of the photo-induced soft phonon dynamics, structural phase transitions, ferroelectricity, and charge density waves.

An alternative \emph{ab-initio} method to track carrier dynamics upon laser excitation is based on real-time time-dependent density functional theory (rt-TDDFT) and Ehrenfest dynamics, and along the EPC allows to include many-body electron-electron and electron-hole interactions\,\cite{akimov13,akimov14,nie14,zheng19,hu22,liu22,guan22}. In combination with molecular dynamics and real-space lattice distortions, the latter method can account for time-resolved self-consistent renormalization of phonon energies and EPC strengths\,\cite{hu22,liu22}, providing a microscopic information on laser-induced structural transitions\,\cite{guan22} and enhanced superconducting properties\,\cite{hu22,liu22}. Since it relies on real-space distortions and supercell approaches to account for electron-lattice interactions, it is numerically challenging for the rt-TDDFT to provide full momentum-resolved analysis on phonon dynamics and, in practice, usually considers only very few coherent optical phonons, such as zone-center $E_{2g}$ and zone-edge $A_1'$ modes in graphene\,\cite{hu22}, and zone-center $A_{1g}$ mode in MoS$_2$\,\cite{liu22}.

On the other hand, the non-thermal renormalization of phonons in the whole Brillouin zone can be acquired from the adiabatic density functional and density functional perturbation theories by constraining the occupation of electronic states to a nonequilibrium distribution (i.e., from cDFT and cDFPT)\,\cite{tangney99,murray05,murray07}, simulating thus laser-excited nonequilibrium states fixed in time, such as population inversion in semiconductors and semimetals. For instance, these methods were used to study photo-induced phonon frequency modifications in tellurium\,\cite{tangney99} and bismuth\,\cite{murray05,murray07}, phase transition in ferroelectrics\,\cite{paillard19} and MoTe$_2$\,\cite{peng20}, as well as structural rippling of hBN monolayer\,\cite{liukun22}. However, the cDFT and cDFPT approaches are time independent and lack information on carrier population dynamics and its temporal evolution towards thermal equilibrium caused by various scattering events.

Here we combine the TDBE and many-body phonon self-energy calculations in order to obtain first-principles information on the time-dependent phonon renrmalization process and electron-phonon scattering channels following laser excitation with a full momentum and frequency resolution. With this methodology we investigate a photoexcited MoS$_2$ monolayer, a prototypical 2D semiconductor with exceptional electronic\,\cite{waldecker19} and optoelectronic\,\cite{qiu13,molinasanchez13,pogna16,bertoni16,mueller18,wood20} properties for which vibrational, electronic, valley, spin, and other degrees of freedom play an active and important role.
Thermalization of carrier distribution function with band and momentum resolution $f_{n\mathbf{k}}$ is obtained as a function of time by means of TDBE where electron-phonon and phonon-phonon scatterings are included\,\cite{caruso21,caruso22,britt22}. Thus acquired distribution functions $f_{n\mathbf{k}}$ are then utilized to construct the time-resolved phonon self-energy $\pi_{\mathbf{q}\nu}(\omega;t)$ and the full nonadiabatic phonon spectral functions $B_{\mathbf{q}\nu}(\omega;t)$, which enables analysis of phonon frequency and linewidth (i.e., relaxation rate) modifications. Characteristic multi-valley landscape of MoS$_2$ electronic states in momentum space permits only selective population dynamics and anisotropic electron-phonon scatterings. Namely, photo-holes in $\mathbf{k}=\mathrm{\Gamma}$ and $\mathbf{k}=\mathrm{K}$ valance valleys promote specifically $\mathbf{q}=\mathrm{\Gamma}$ and $\mathbf{q}=\mathrm{K}$, while photo-electrons in $\mathbf{k}=\mathrm{K}$ and $\mathbf{k}=\mathrm{Q}$ conduction valleys promote dominantly $\mathbf{q}=\mathrm{\Gamma}$ and $\mathbf{q}=\mathrm{M}$ electron-phonon scatterings of optical and acoustic phonons. This in turn influences considerably phonon frequency and linewidths, and results in remarkable anisotropic nonequilibrium phonon softening and dynamical Kohn anomalies at aforesaid phonon momenta. For instance, large dynamical Kohn anomaly of the $A_{1g}$ optical mode appears close to $\mathbf{q}=\mathrm{\Gamma}$, surpassing the strength of the corresponding phonon anomaly in equilibrium state of doped MoS$_2$ samples\,\cite{sohier19,novko2020a,eiguren20}. Also, sizeable phonon softening is induced for the longitudinal acoustic (LA) phonon at $\mathbf{q}=\mathrm{M}$, which is considered relevant for the appearance of the superconductivity\,\cite{fu17} and CDW\,\cite{subhan21} in doped MoS$_2$. Importantly, we show that overall phonon scattering rate is significantly increased in nonequilibrium, opening a possibility for enhancing the total EPC strength. These findings demonstrate that photo-induced nonequilibrium state is a promising route for tailoring vibrational properties of quantum matter, especially for MoS$_2$ where phonons play a primary role in the emergence of novel quantum phenomena, such as in exciton dynamics\,\cite{carvalho15,reichardt20,chan23,bange23} as well as formation of Holstein polaron\,\cite{kang18}, CDW\,\cite{subhan21} and superconductivity\,\cite{lu15,costanzo16,fu17,piatti18}.

\section{Theoretical methods}
The dynamics of non-thermal electron-lattice system in the TDBE is described by modifications of electron and phonon occupation functions $f_{n\mathbf{k}}(t)$ and $n_{\mathbf{q} \nu}(t)$, while electron and phonon energies, as well as the corresponding coupling functions are unaltered and fixed to their equilibrium values. The time evolution of $f_{n\mathbf{k}}(t)$ and $n_{\mathbf{q} \nu}(t)$ is dictated by electron-phonon and phonon-phonon scattering processes and can be described with the following coupled integro-differential equations\,\cite{caruso21,caruso22}:
\begin{align}
\label{eq:bte1}  {\partial_t f_{n\mathbf{k}}(t)}     &= \Gamma^{\rm ep}_{n\mathbf{k}}[f_{n\mathbf{k}}(t), n_{\mathbf{q}\nu}(t)], \\
\label{eq:bte2}  {\partial_t n_{\mathbf{q} \nu}(t)}  &= \Gamma^{\rm pe}_{{\mathbf{q}\nu}}[f_{n\mathbf{k}}(t), n_{\mathbf{q}\nu}(t)]  + \Gamma^{\rm pp}_{\mathbf{q}\nu}[n_{\mathbf{q}\nu}(t)],
\end{align}
where $\partial_t = \partial / \partial t$, while $\Gamma_{n\mathbf{k}}$ and $\Gamma_{{\mathbf{q}\nu}}$ denote the collision integrals for electrons and phonons, for which electron-phonon, phonon-electron, and anharmonic phonon-phonon scatterings are accounted for. The corresponding expressions for the collision rates are derived from the standard first-order Fermi's golden rule and the explicit forms can be found in Refs.\,\citenum{caruso21,britt22,caruso22}. The time dependence of the collision integrals is explicitly accounted for in the calculations by reevaluating $\Gamma^{\rm ep}_{n\mathbf{k}}$ and $\Gamma^{\rm pe}_{{\mathbf{q}\nu}}$ at each time step of the time propagation with new occupation functions. The phonon-phonon collision integral is treated with the relaxation-time approximation.
Electron-electron scattering channel governs the thermalization of photoexcited carriers when $f_{n\mathbf{k}}(t)$ deviates significantly from the equilibrium distribution.
Here we consider exclusively electronic excited state characterized by a weak deviation from a Fermi-Dirac function. In this regime, electron-electron scattering plays only a minor role in the carrier dynamics and it is therefore neglected.

At thermal equilibrium, $f_{n\mathbf{k}}$ and $n_{\mathbf{q}\nu}$
are time independent and they coincide with the Fermi-Dirac and the
Bose-Einstein occupations: 
$
f^{\rm FD}_{n\mathbf{k}} = \left[ e^{ {(\varepsilon_{n\mathbf{k}}-\varepsilon_{\rm F})}/{k_{\rm B}T}} + 1 \right]^{-1},
n^{\rm BE}_{\mathbf{q}\nu} =\left[ e^{ \hbar\omega_{\mathbf{q}\nu}/{k_{\rm B}T}} - 1 \right]^{-1}.
$
Here, $\varepsilon_{\rm F}$  is the Fermi energy, 
$\varepsilon_{n\mathbf{k}}$ is the single-particle energy of a Bloch electron, and ~$\hbar\omega_{\mathbf{q}\nu}$ the phonon energy. The initial photoexcited concentration of electrons is taken to be $n=10^{14}$\,cm$^{-2}$. The corresponding initial photo-holes and photo-electrons are defined with two separate chemical potentials and high carrier temperature, while keeping in mind the conservation of carrier number. Namely, we define $f_{n\mathbf{k}}(t=0)=f^{\rm FD}_{n\mathbf{k}}(\mu_{\rm e/h},T^0_{e})$, with $\mu_{\rm e}$ ($\mu_{\rm h}$) being the electron (hole) chemical potential, $T^0_{\rm e}=2000$\,K, while phonon temperature is set to $T^0_{\rm p}=100$\,K. Equations \eqref{eq:bte1} and \eqref{eq:bte2} are solved by time-stepping the derivatives with small time step of 1\,fs up to 40\,ps.

In order to have a full time-dependent information of electron-phonon dynamics an important step forward is to update the phonon frequencies coming from the modified electron occupation functions $f_{n\mathbf{k}}(t)$. This can be calculated by means of the time-resolved phonon spectral function defined as
\begin{equation}
B_{\mathbf{q}\nu}(\omega;t)=-\frac{1}{\pi}\mathrm{Im}\left[ \frac{2\omega_{\mathbf{q}\nu}}{\omega^2-\omega_{\mathbf{q}\nu}^2-2\omega_{\mathbf{q}\nu}\widetilde{\pi}_{\mathbf{q}\nu}(\omega;t)} \right].
\label{eq:bspec}
\end{equation}
The crucial ingredient to the above expression is the time-resolved NA phonon self-energy\,\cite{giustino17} defined as $\widetilde{\pi}_{\mathbf{q}\nu}(\omega;t)=\pi_{\mathbf{q}\nu}(\omega;t)-\pi_{\mathbf{q}\nu}(0)$, where the adiabatic part at $t\rightarrow -\infty$, i.e., $\pi_{\mathbf{q}\nu}(0)$, is subtracted, and where
\begin{equation}
\pi_{\mathbf{q}\nu}(\omega;t)=\sum_{\mathbf{k}nm}\left| g_{\nu}^{nm}(\mathbf{k},\mathbf{q}) \right|^2\frac{f_{n\mathbf{k}}(t)-f_{m\mathbf{k+q}}(t)}{\omega+\varepsilon_{n\mathbf{k}}-\varepsilon_{m\mathbf{k+q}}+i\eta}.
\label{eq:phonse}
\end{equation}
The electron-phonon matrix elements are denoted by $g_{\nu}^{nm}(\mathbf{k},\mathbf{q})$ and $\eta$ is an infinitesimal parameter. In our approach, the electron occupation functions $f_{n\mathbf{k}}(t)$ entering the phonon spectral function \eqref{eq:bspec} and phonon self-energy \eqref{eq:phonse} are no longer Fermi-Dirac distributions as in the standard thermal case\,\cite{giustino17}, but are extracted from TDBE Eq.\,\eqref{eq:bte1} and, therefore, represent nonequilibrium occupations and redistribution of charge carriers at each time instant after the laser excitation. 
The corresponding photo-induced renormalization of phonon frequency and modifications to the phonon linewidth (relaxation rate) can be tracked as a function of time by means of the following expressions\,\cite{giustino17}
\begin{align}
\Omega^2_{\mathbf{q}\nu}(t)&=\omega^2_{\mathbf{q}\nu}+2\omega_{\mathbf{q}\nu}{\rm Re}\,\widetilde{\pi}_{\mathbf{q}\nu}(\Omega_{\mathbf{q}\nu}(t);t), \label{eq:freqna}\\
\gamma_{\mathbf{q}\nu}(t)&=-{\rm Im}\,\widetilde{\pi}_{\mathbf{q}\nu}(\Omega_{\mathbf{q}\nu}(t);t). \label{eq:linena}
\end{align}
A similar idea was adopted recently in Refs.\,\citenum{seiler21,britt22} where occupation functions as obtained from TDBE were introduced into dynamic structure factors to complement ultrafast electron diffraction experiments.

\begin{figure}[!t]
    \centering
    \includegraphics[width=0.45\textwidth]{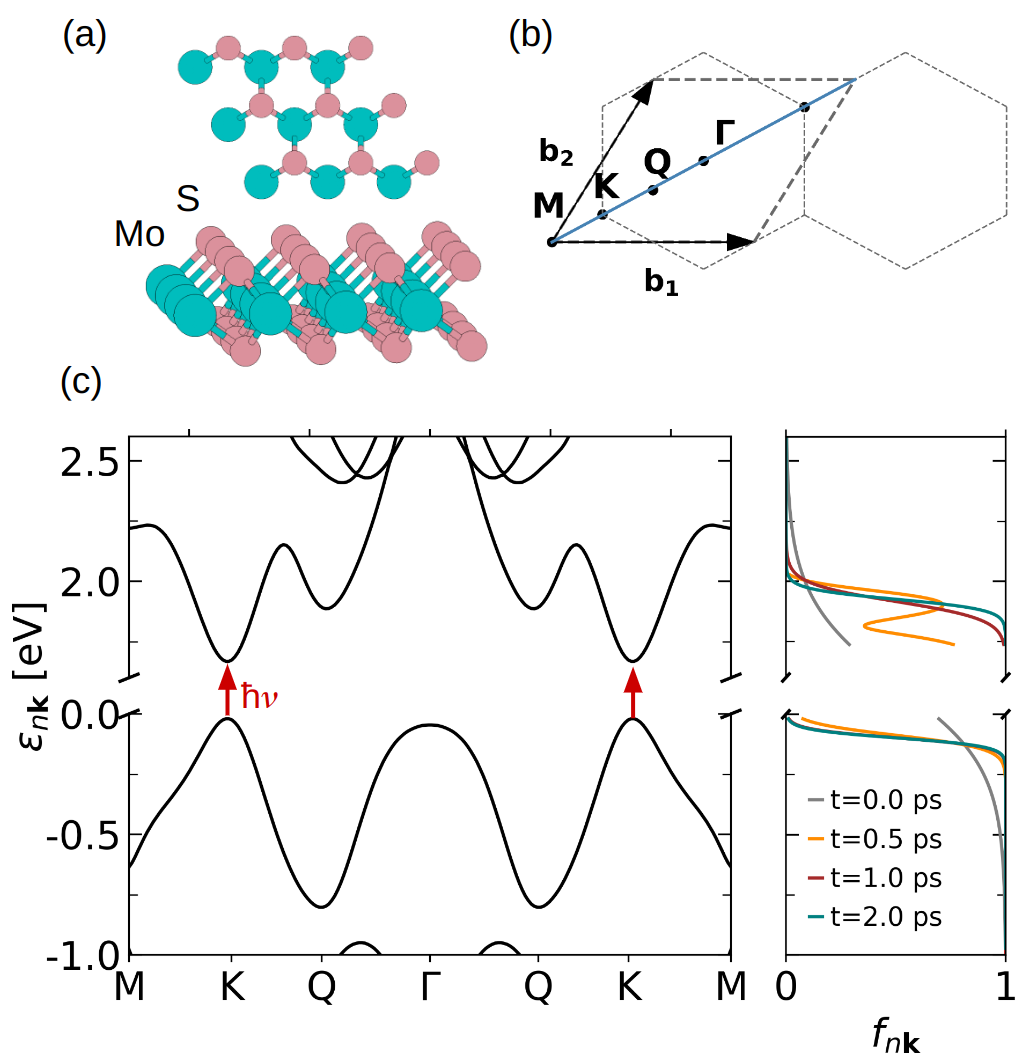}
    \caption{(a) Crystal structure of MoS$_2$ single layer. (b) Corresponding Brillouin zone and high-symmetry points. (c) Electronic band structure of MoS$_2$ with time evolution of the photoexcited occupation functions as obtained from the time-dependent Boltzmann equations.}
    \label{fig:fig1}
\end{figure}

\begin{figure*}[!t]
    \centering
    \includegraphics[width=\textwidth]{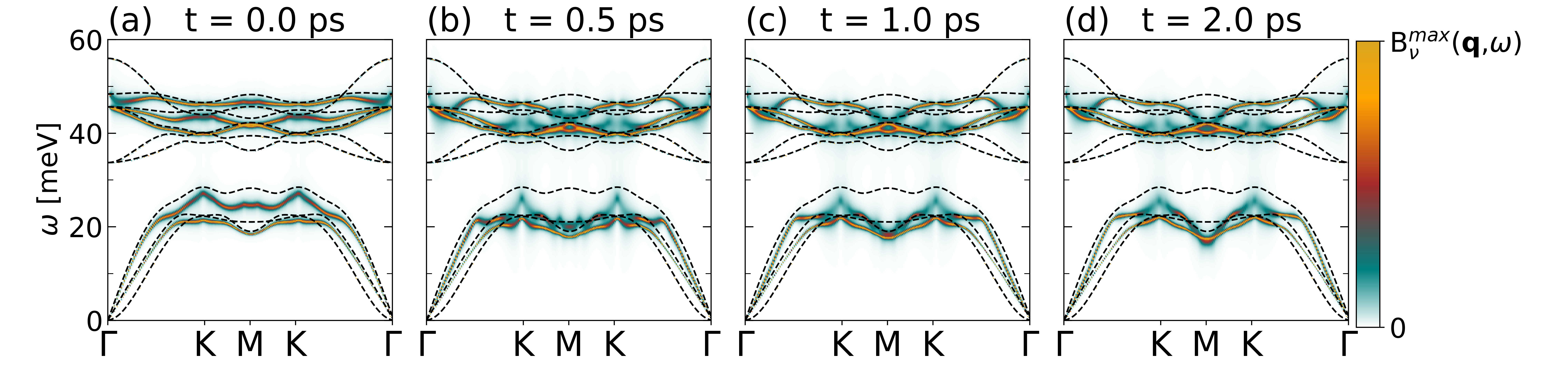}
    \caption{Phonon spectral functions $B_{\nu}(\mathbf{q},\omega)$ of photoexcited MoS$_2$ shown along high-symmetry points in the first Brillouin zone and for several time frames after excitation: (a) $t=0$\,ps, (b) $t=0.5$\,ps, (c) $t=1$\,ps, and (d) $t=2$\,ps. The dashed lines are phonon dispersions for pristine equilibrium MoS$_2$ as obtained from the adiabatic DFPT. Significant photo-induced phonon broadening enhancements and dynamical Kohn anomalies can be observed around the high-symmetry points both for optical and acoustic branches.}
    \label{fig:fig2}
\end{figure*}

We also define the spectral representation of the phonon scattering rate $\gamma F(\omega)$ in order to quantify the modifications to the electron-phonon scattering channels relevant for phonon dynamics and to discuss possible enhancements to the total EPC strength upon the laser excitation
\begin{align}
\gamma F(\omega;t)=\sum_{\mathbf{q}\nu}\gamma_{\mathbf{q}\nu}(t)\delta\left(\omega-\Omega_{\mathbf{q}\nu}(t)\right),
\label{eq:gf}
\end{align}
where $\delta(x)$ is the Dirac delta function. Note that the above spectral function is defined in a similar manner as the Eliashberg function or electron-phonon spectral function $\alpha^2 F(\omega)$\,\cite{allen74}. The cumulative scattering rate of phonons can then be written as
\begin{align}
\gamma(\omega;t)=\int_{0}^{\omega}d\omega' \gamma F(\omega';t),
\label{eq:gw}
\end{align}
while the total phonon scattering rate is $\gamma(\omega\rightarrow \infty;t)$.

All of the above equations and the corresponding input parameters are calculated in this work by means of DFPT\,\cite{baroni01} and Wannier interpolation\,\cite{wan90} of EPC matrix elements $g_{\nu}^{nm}$\,\cite{epw}.

We use \textsc{Quantum ESPRESSO}~\cite{Giannozzi2009, Giannozzi2017, Giannozzi2020} for the DFT calculations, and for EPC we use the EPW code~\cite{Giustino2007a, Noffsinger2010, Ponce2016}. All calculations are performed with the norm-conserving Perdew-Burke-Ernzerhof pseudopotential with a kinetic energy cutoff of 120\,Ry. The lattice constant is set to the value of 3.188\,\AA, while the neighboring MoS$_2$ sheets are separated by 12.7\,\AA. The self-consistent electron density calculation is done on a $20 \times 20\times 1$ k-point grid and the phonon calculation on a $6 \times 6 \times 1$ q-point grid. Both are done with the equilibrium electron occupation functions for pristine MoS$_2$, with the valence band occupied and the conduction band unoccupied. To interpolate electron-phonon quantities, we use 11 maximally localized Wannier functions~\cite{Marzari2012} with the initial projections of d-orbitals on the Mo sites and p-orbitals on the S atom sites. All electronic structure parameters for the self-consistent cycle and the Wannierization match the ones used to solve the TDBE. The fine sampling of the Brillouin zone for the electron-phonon interpolation is done on a $200 \times 200\times 1$ grid. The fine q-point grid is always extracted among these 40000 points, whether it is a path in q-space or a full Brillouin zone calculation. Smearing in the EPW calculation is set to 40\,meV.

\section{Results and discussion}

Figure \ref{fig:fig1} depicts the crystal structure of single-layer MoS$_2$, the corresponding Brillouin zone and electronic band structure along high-symmetry points. MoS$_2$ is a semiconducting TMD with a direct band gap at the K point of the Brillouin zone, and interesting multi-valley topology of valence and conduction bands. The latter is considered to be instrumental for various physical phenomena in MoS$_2$, such as enhanced EPC of the $A_{1g}$ phonon mode as observed with Raman spectroscopy\,\cite{sohier19}, exciton-phonon coupling\,\cite{chan23,bange23}, anisotropic electron-phonon scattering following laser excitation\,\cite{britt22}, and multi-valley superconductivity\,\cite{lu15,costanzo16,fu17,piatti18}. As it will be shown below, the rich phase space for electron-phonon scatterings come from the $\Gamma$ and K valleys in the valence band and the K and Q valleys in the conduction band.

Additionally, in Fig.\,\ref{fig:fig1}(c) we show occupation functions $f_{n\mathbf{k}}(t)$ for several time instants up to $t=2$\,ps as obtained from the solution of the TDBE \eqref{eq:bte1}-\eqref{eq:bte2}. Initial hot distribution of photo-holes and photo-electrons (grey lines) thermalize to smaller effective electron temperatures (about 180\,K) with different time scales. Namely, holes are almost thermalized at $t=0.5$\,ps, while it takes around 2\,ps for excited electrons to equilibrate. The reason for this is the difference in the phase space of valence and conduction bands, where the Q valley acts as a sort of bottleneck for electron-phonon scattering in the conduction band and slows down the process (see the accumulated charge for $t=0.5$\,ps that forms non-Fermi-Dirac distribution). On the other hand, the obained thermalization time of a nonequilibrium phonon distribution is somewhere between 5 and 10\,ps\,\cite{caruso21}. Note that while at the initial step and after thermalization time the population of electrons and phonons are well described with quasi-equilibrium distribution functions having corresponding temperatures, the TDBE allows for large deviations from the equilibrium in the between, as it is for instance the case for the time period between 0.1 and 1\,ps for electrons, and between 0.1 and 5\,ps for phonons.

The impact of the nonequilibrium electron and hole distributions on phonons is shown in Fig.\,\ref{fig:fig2} where we report the full phonon spectral functions for several time delays. Considerable and time-varying nonequilibrium phonon renormalizations are observed for both acoustic and optical branches, especially around the high-symmetry points of the BZ. In addition, these renormalizations are accompanied by a remarkable enhancement of the phonon broadenings. Particularly strong modifications of phonon frequency is observed for $A_{1g}$ optical phonon at $\mathbf{q}=\Gamma$, where a much larger dynamical Kohn anomaly is formed by excited carriers compared to the case of doped MoS$_2$ in equilibrium state\,\cite{novko2020a}. 
Kohn anomalies are distinct softenings of phonon dispersion coming from singularities in static phonon self-energy $\pi_{\mathbf{q}\nu}(0)$, which in turn come from highly anisotropic electron-phonon matrix elements $g_{\nu}^{nm}(\mathbf{k},\mathbf{q})$ as well as from the intense and anisotropic density of states of electron-hole pair excitations\,\cite{giustino17}. If the Kohn anomaly calculated in nonadiabatic regime [i.e., with $\pi_{\mathbf{q}\nu}(\omega)$] is different from the static one, it is usually dubbed dynamical.
The dynamical Kohn anomaly of optical phonons at $\mathbf{q}=\Gamma$ was reported for various single-layer and bulk materials (such as graphene, hole-doped diamond, MgB$_2$, and doped TMDs) in thermal equilibrium\,\cite{lazzeri06,caruso17,novko2018a,novko2020a,berges22,girotto23}. Here 
we illustrate to which extent a photoexcited carrier distribution  can provide a route to trigger and control the emergence of Kohn anomalies over transient timescales.
The $A_{1g}$ mode was shown to be significantly coupled to electrons once multiple valleys are occupied\,\cite{sohier19}, as it is the case here for photoexcited MoS$_2$. The same phonon mode is as well quite affected at the M point, while energetically lower $E_{2g}$ optical mode is softened and broadened at the K point. Also, both longitudinal and transverse acoustic modes are modified, especially around the M and K points. Softening of the LA mode at M point particularly stands out, where at time $t=2$\,ps its frequency is decreased by about 10\,meV compared to the equilibrium value. Since the LA mode at $\mathbf{q}=\mathrm{M}$ is instrumental for phonon-mediated superconductivity\,\cite{fu17} as well as the CDW formation\,\cite{subhan21}, these results suggest a possibility to tune these ordered states by laser excitations.

\begin{figure*}[!ht]
    \centering
    \includegraphics[width=\textwidth]{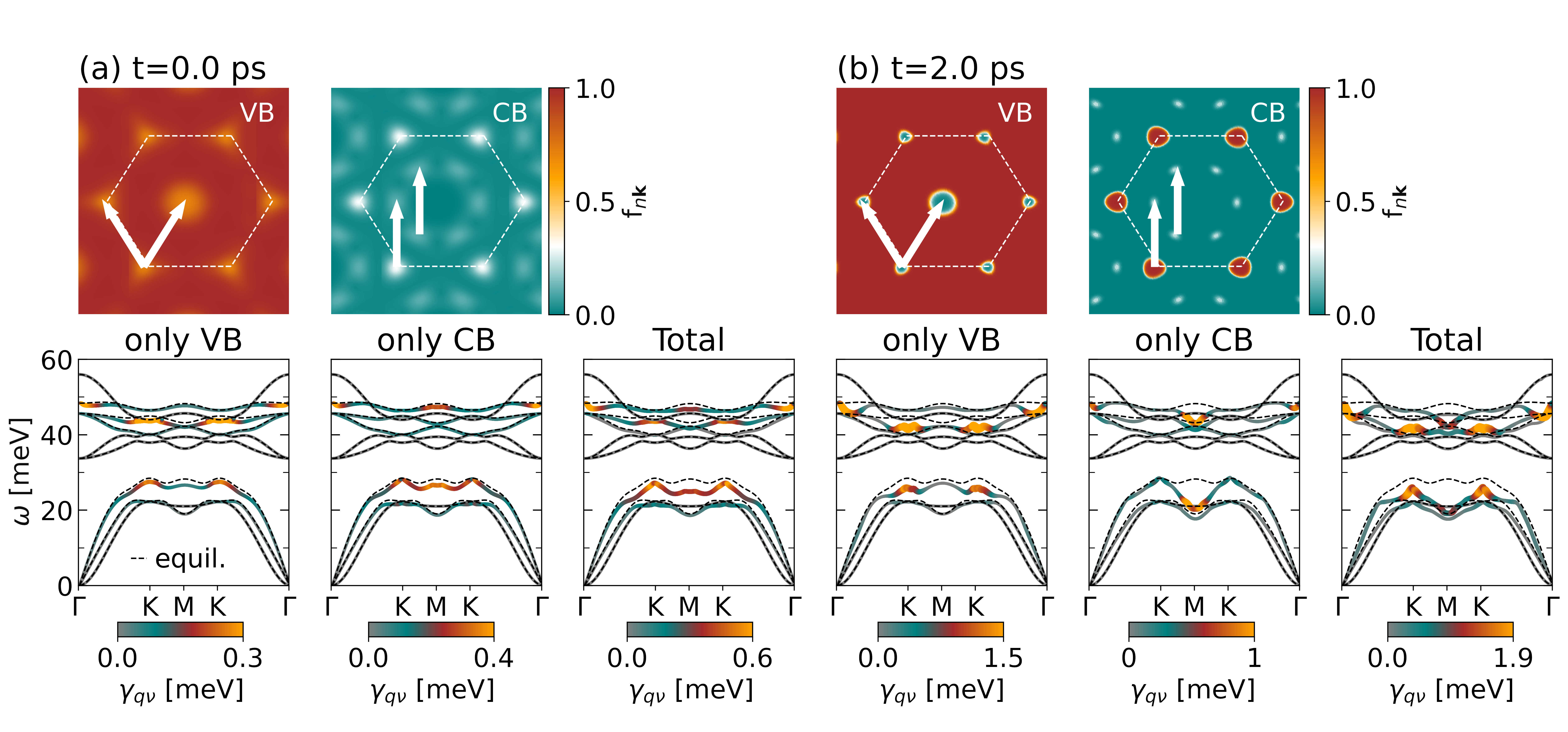}
    \caption{Nonequilibrium electron-phonon scattering channels coming from valence- (VB) and conduction-band (CB) distributions. The VB- and CB-resolved contributions are presented for two time instants (a) $t=0\,\mathrm{ps}$ and (b) $t=2\,\mathrm{ps}$. Upper panels depict the momentum-resolved occupation functions $f_{n\mathbf{k}}(t)$ for the CB and VB within the first Brillouin zone. White arrows represent the dominating inter-valley scatterings channels. The lower panels show the VB- and CB-resolved contributions to the phonon dispersion and linewidth $\gamma_{\mathbf{q}\nu}$. In addition, the overall result of dynamical phonon renormalization, coming from both VB and CB nonequilibrium channels, is shown. The dashed lines again show the phonon dispersions for pristine equilibrium MoS$_2$ as obtained from the adiabatic DFPT}
    \label{fig:fig3}
\end{figure*}

\begin{figure}[!t]
    \centering
    \includegraphics[width=0.48\textwidth]{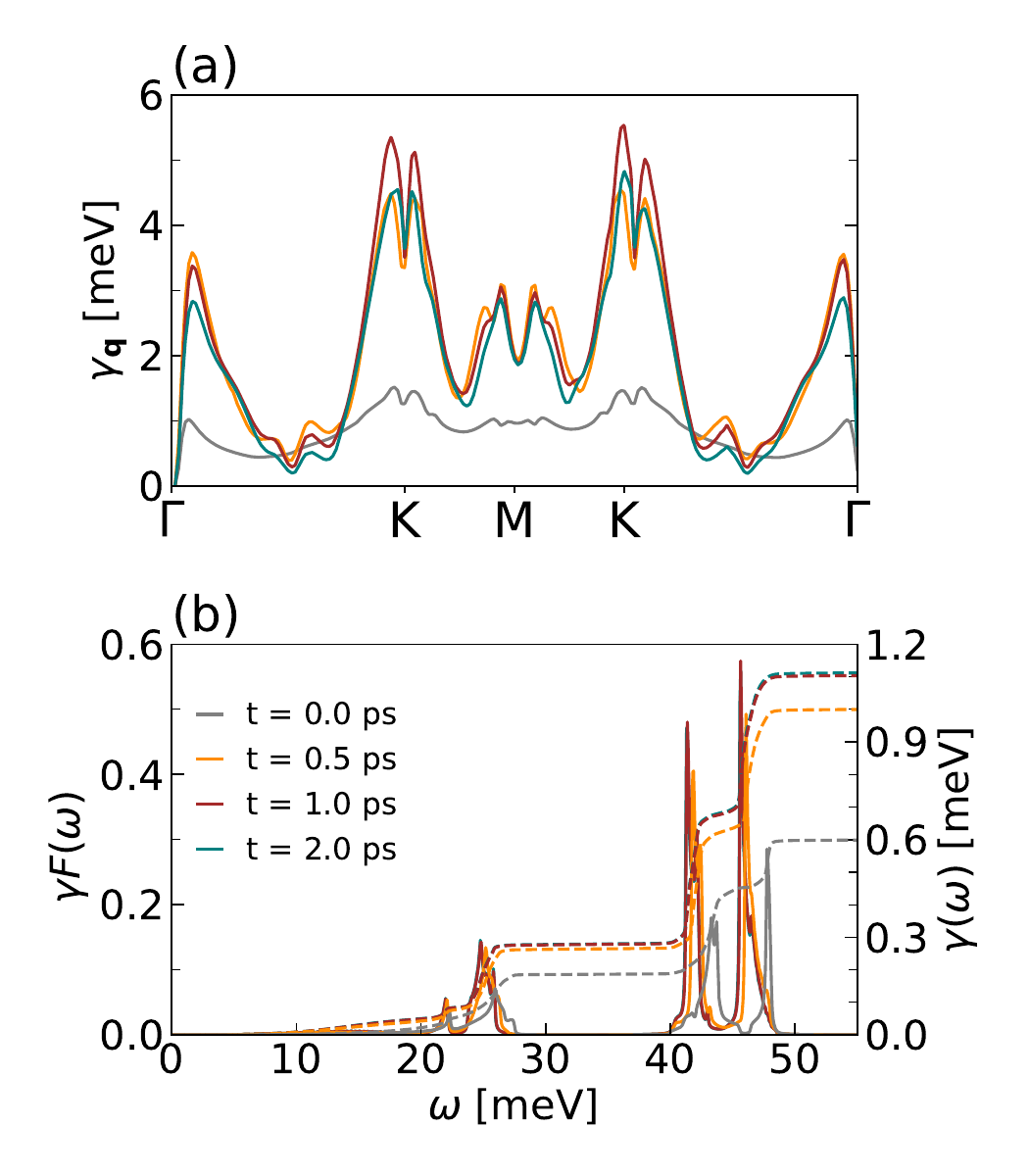}
    \caption{(a) Total phonon scattering rate (summed over all branches) along the high-symmetry points and as a function of time. (b) Spectral representation of phonon scattering rates $\gamma F(\omega)$ (phonon density of states weighted with phonon linewidth contributions) as a function of frequency $\omega$ shown for several time delays. Right axis shows the cumulative scattering rate $\gamma(\omega)$.}
    \label{fig:fig4}
\end{figure}

In the following we study the phase space arguments and analyze the electron-phonon scattering events out of equilibrium that lead to these remarkable phonon renormalizations and phonon linewidth enhancements. In the top panels of Figs.\,\ref{fig:fig3}(a) and \ref{fig:fig3}(b) we show the momentum-resolved occupation functions $f_{n\mathbf{k}}(t)$ of valence and conduction bands for $t=0$ and $2$\,ps, respectively. The corresponding bottom panels show the contributions to the phonon dispersions and linewidths coming only from the electron-phonon scatterings in the valence and conduction bands. Both time frames are characterized by the similar phase space for the valence and conduction bands, i.e., depopulated $\mathbf{k}=\Gamma$ and $\mathbf{k}=\mathrm{K}$ valleys in the valence band and populated $\mathbf{k}=\mathrm{K}$ and $\mathbf{k}=\mathrm{Q}$ valleys in the conduction band. Such occupations promote $\mathbf{q}=\Gamma$ and $\mathbf{q}=\mathrm{K}$ electron-phonon scatterings in the former, while $\mathbf{q}=\Gamma$, $\mathbf{q}=\mathrm{K}$ and $\mathbf{q}=\mathrm{M}$ electron-phonon scatterings within the latter band. Consequently, these intra- and inter-valley scatterings in the valence band lead to nonequilibrium renormalization of the optical $A_{1g}$ and $E_{2g}$ phonons at $\Gamma$ and K points, respectively, as well as of the LA mode at the K point. Similarly, the scattering channels in the conduction band produce phonon softenings of the optical $A_{1g}$ phonon at $\Gamma$ and M points, and the LA mode at the K and M points.
As the initially hot distribution of electrons and holes at $t=0$\,ps is distributed into more sharp occupations at the top and bottom of valence and conduction bands at $t=2$\,ps, some alterations of the phonon bands and broadenings become more pronounced while others are reduced. For instance, as the K valley becomes more populated, while the occupation in the Q valley is reduced in the conduction band, the corresponding modifications of the $\mathbf{q}=\mathrm{K}$ ($\mathbf{q}=\mathrm{M}$) phonons coming only from conduction-band scatterings are reduced (enhanced).

Femtosecond electron diffraction experiment revealed that scatterings in multi-layer MoTe$_2$ are dominated by the zone-center $A_{1g}$ and $E_{2g}$ optical phonons, as well as by the LA phonons at the M point of the BZ\,\cite{krishnamoorthy19}. Momentum-resolved picture of the energy transfer between excited electrons and phonons in thin bulk-like films of WSe$_2$ reveals importance of inter-valley scattering between two Q points followed by emission of the acoustic M-point phonons\,\cite{bertoni16,waldecker17}.
On the other hand, coherent phonon dynamics extracted from femtosecond pump-probe spectroscopy had shown that ultrafast intervalley scattering in monolayer MoSe$_2$ is dictated dominantly by the LA phonons, but at K point of the BZ\,\cite{bae22}. Multi-layer films of MoTe$_2$ and WSe$_2$ are indirect-band-gap semiconductors, and are therefore characterized by different scattering phase space for excited carriers compared to monolayer MoSe$_2$, which has a direct band gap at the K point. Consequently, monolayer and thin films of TMDs have distinct ultrafast phonon dynamics, coming from different shapes of conduction and valence valleys. 
For instance, in bulk-like semiconducting TMDs the Q conduction valley has lower energy than the K valley, and therefore the $\mathrm{Q}\leftrightarrow\mathrm{Q'}$ inter-valley scatterings are dominant with emission of the $\mathbf{q}=\mathrm{M}$ phonons. For corresponding single layers the situation is different, where the K conduction valley has the lowest energy, and more dominant are the $\mathrm{K}\leftrightarrow\mathrm{K'}$ scatterings (actually in the both conduction and valence bands) and the concomitant $\mathbf{q}=\mathrm{K}$ phonon emission.
This opens many possibilities to tailor ultrafast phonon scatterings, e.g., by strain, pressure, doping, and other techniques that can significantly alter the energy positions of valleys\,\cite{blundo21}.

To further demonstrate the implications of nonequilibrium carrier distribution on phonon dynamics, we show in Fig.\,\ref{fig:fig4} the total phonon scattering rate (summed over all branches) as a function of time along the high-symmetry points and spectral representation of phonon scattering rates $\gamma F(\omega)$ with the corresponding cumulative scattering rate $\gamma(\omega)$ [see Eqs.\,\eqref{eq:gf} and \eqref{eq:gw}]. The momentum resolved scattering rate $\gamma_{\mathbf{q}}$ reveals important role of the $\mathbf{q}=\Gamma$ and $\mathbf{q}=\mathrm{K}$ phonons and their dominance in the overall phonon relaxation dynamics. Importance of these specific phonon modes are in line with the ultrafast coherent phonon dynamics in single-layer MoSe$_2$\,\cite{bae22}, while anisotropic phonon response is in accordance with recent results obtained with ultrafast electron diffraction spectroscopy in MoS$_2$ monolayer\,\cite{britt22}. Certain discrepancies in timescales between experiments and theoretical results obtained here could be attributed to the screening of the matrix element induced by the substrate\,\cite{britt22}. Note that the recent theoretical study based on nonequilibrium Green's functions also obtained that the $\Gamma$- and K-point phonons are dominantly involved in nonequilibrium carrier dynamics in monolayer MoS$_2$\,\cite{perfetto23}. Interestingly, they also show that the optical phonons participate more in the relaxation dynamics compared to the acoustic modes.

The obtained relaxation rates around $\Gamma$, K, as well as M points are also significantly increased in time, where values at $t=2$\,ps are $4-5$ times larger compared to the corresponding values at the initial time. The frequency-resolved scattering rate is presented in Fig.\,\ref{fig:fig4}(b) via $\gamma F(\omega)$, where it shown that nonequilibrium phonon scattering rates is larger for optical modes above 40\,meV. The results for cumulative rate $\gamma(\omega)$ confirms the gradual increase of the total rate as a function of time. This result seems surprising at first, since one expects that vibrational energy exchange rate decreases towards the thermalization instant, i.e., when the effective electronic temperature is decreased\,\cite{waldecker17}. Nevertheless, in some cases (like nickel and platinum as well as photo-doped MoS$_2$ as shown here) due to specific Fermi surface and density of states, the opposite is possible\,\cite{lin08}.

\begin{figure}[!t]
    \centering
    \includegraphics[width=0.48\textwidth]{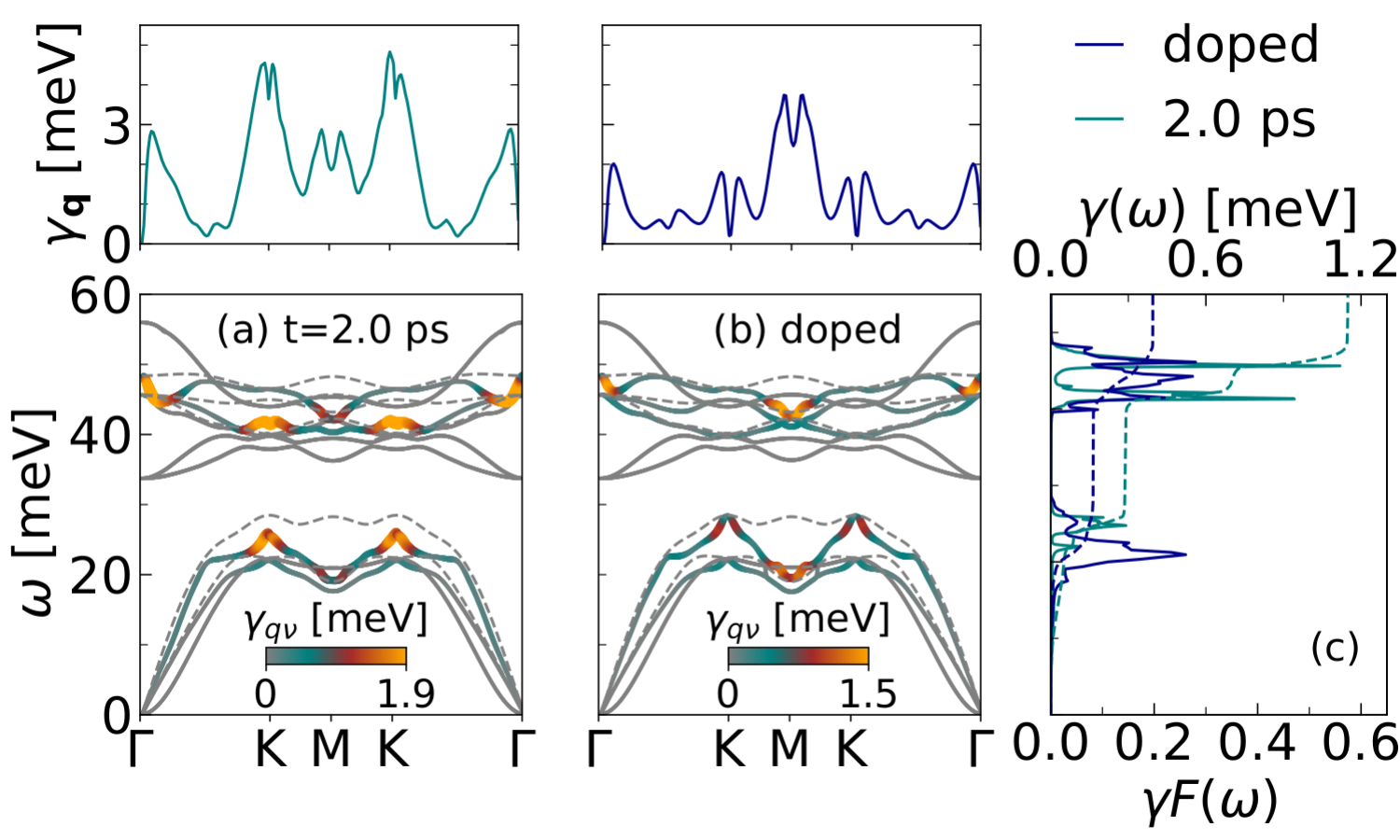}
    \caption{(a) Total and mode-resolved nonequilibrium phonon scattering rates along the high-symmetry points for $t=2$\,ps. (b) Same as (a) but for the case of electron-doped equilibrium MoS$_2$ for the effective carrier concentration $n_{\rm eff}=2\times 10^{14}$\,cm$^{-2}$ that matches the photoexcited carrier (i.e., both photo-hole and photo-electron) density at $t=2$\,ps. (c) Spectral representation of phonon scattering rates $\gamma F(\omega)$ as a function of frequency $\omega$ and the cumulative scattering rate $\gamma(\omega)$ for the photo-doped and doped cases presented in (a) and (b) panels.}
    \label{fig:fig5}
\end{figure}

In addition, we want to explore the difference in phonon dynamics between the photo-doped (i.e., photoexcited) scenario investigated here and the standard case of the electron-doped material via field-effect techniques or atom adsorption (i.e., via dopants). Figure \ref{fig:fig5} compares the total and mode-resolved phonon scattering rates along the high-symmetry points for photoexcited MoS$_2$ at $t=2$\,ps [panel (a)] and MoS$_2$ doped with electron carrier concentration of $n_{\rm eff}=2\times 10^{14}$\,cm$^{-2}$, which corresponds to the effective photo-induced carrier (both electron and hole) density at $t=2$\,ps [panel (b)]. For the electron-doped case, the active phase space for electron-phonon scatterings consists only of the K and Q valleys in the conduction band. This in turn promotes dominantly intra-valley $\mathbf{q}=\Gamma$ and inter-valley $\mathbf{q}=\mathrm{M}$ scatterings and consequently phonon renormalizations and broadening around these symmetry points. Furthermore, in Fig.\,\ref{fig:fig5}(c) we show spectral representation of phonon scattering rates $\gamma F(\omega)$ and the cumulative scattering rate $\gamma(\omega)$ for these two cases, where it is clear that photo-doping induces larger and richer phonon-electron scatterings, since the corresponding scattering phase space includes both photo-holes and photo-electrons, i.e., both valence and conduction valleys.

Overall, the present results clearly demonstrate a notable increase of phonon-electron scattering rate with photoexcitation, which points to the potential of enhancing the total EPC strength out of equilibrium\,\cite{liu22} as well as inducing and modifying the concomitant superconducting properties. Note that in order to have a more conclusive answer to the intriguing physical problem of photo-induced superconductivity, one would need to go beyond the present consideration and adopt more rigorous time-dependent methodology of superconductivity, such as nonequilibrium Green's function techniques\,\cite{giannetti16,sentef16,murakami17,sentef17,kemper17}.

As a final remark, we want to note that, besides on the phonon dynamics as studied here, the laser-induced nonequilibrium carrier and phonon distributions could potentially have a significant impact on the electron dynamics, such as ultrafast band renormalizations. Within the present theoretical framework, these ultrafast features could be captured by updating the electronic energies and linewidths with electron self-energy due to EPC, which are in turn based on the TDBE results. This future direction might provide some interesting microscopic insights on the observed ultrafast band gap renormalizations in TMDs\,\cite{pogna16,wood20} and other layered materials\,\cite{mor17}, as well as on the intriguing Floquet physics, such as phonon-driven Floquet matter\,\cite{hubener18} and ultrafast quasiparticle dressing by light\,\cite{reutzel20}.

\section{Conclusions}
We have explored the phonon relaxation pathways and the ensuing nonequilibrium phonon renormalization in the photoexcited MoS$_2$ monolayer by combining the \emph{ab-initio} time-dependent Boltzmann equation and the phonon self-energy calculations. Our findings show how population and depopulation of conduction and valence valleys promote anisotropic electron-phonon scatterings and triggers strong Kohn anomalies, softenings, and increase of relaxation rate for strongly-coupled optical and acoustic phonon modes. Nonequilibrium of the electronic energy levels induces strong Kohn anomaly of the $E_{2g}$ mode close to the center and edge of the Brillouin zone, and strongly softens the longitudinal acoustic phonon at the M point. In accordance to the recent ultrafast experiments, our momentum-resolved analysis demonstrates that the $\mathbf{q}=\Gamma$ and $\mathbf{q}=\mathrm{K}$ phonon modes play a key role in intra- and inter-valley scattering channels and they are thus characterized by large relaxation rates. It is also shown that as the effective electron temperature decreases, i.e., as the photo-holes and photo-electrons are scattered towards the top and bottom of valence and conduction valleys, the overall phonon relaxation rate is significantly enhanced.
The richness of the phase space for the photo-carriers and the corresponding impact on phonon dynamics was further demonstrated in comparison to the electron-doped MoS$_2$ in equilibrium, where, instead of K-point, M-point modes are ruling the phonon relaxation and renormalization, and where phonon relaxation rates have less intensity. In general, we believe that present results and methodology might be instrumental for gaining crucial microscopic insights of photoexcited states in multi-valley systems, such as transition metal dichalcogenides, as well as discovering new photo-induced ordered phases (e.g., superconductivity and charge density waves) and structural transformations in condensed matter.

\begin{acknowledgement}
Useful discussions with Jan Berges, Samuel Ponc\'e, Yiming Pan are gratefully acknowledged. We acknowledge financial support from the Croatian Science Foundation (Grant no. UIP-2019-04-6869) and from the European Regional Development Fund for the ``Center of Excellence for Advanced Materials and Sensing Devices'' (Grant No. KK.01.1.1.01.0001).
F.C. acknowledges funding from the Deutsche Forschungsgemeinschaft Grant No. 443988403 and 499426961.
\end{acknowledgement}

\bibliography{ref}

\end{document}